Ferromagnetic III-Mn-V semiconductors, such as GaMnAs, InMnAs and heterostructures based on them, have revealed many interesting physical properties and spintronic device possibilities. [1, 2] The InAs/GaSb-based heterostructure system has, due to the unique band alignment, the additional advantage of spatially separating electrons and holes, which permits optical and electrical tuning of ferromagnetism. [3, 4] Furthermore, these materials and heterostructures are strong candidates for active components of spintronic devices due to the high mobility of InAs and their extensive applications as infrared sources and detectors. While GaMnAs and InMnAs exhibit well-defined hysteresis loops in the ferromagnetic phase, such clear behavior has been elusive in GaMnSb, in which $T_C$ is low (25 K) and the hysteresis loops are very small. [5]

Driven by the ultimate goal of spintronic device applications at room temperature, extensive experimental and theoretical studies have been performed on various materials and structures. Much of this work has been directed at the design of materials and structures on the atomic scale, so that $T_C$ can be raised to room temperature or above. The highest $T_C$ observed in the III-V materials so far is 110 K in GaMnAs. [6, 7] Theoretical calculations have shown that in the Zener model spin-orbit interaction and the density of states at the Fermi energy are important parameters in determining $T_C$. [8] These calculations, which focused on conventional alloys containing randomly distributed Mn in the host materials, predict room temperature ferromagnetism in GaMnN and ZnMnO. These materials have been investigated experimentally, [9, 10] and ferromagnetism at high temperatures in crystalline ZnCoO has been reported. However, epitaxial growth of both ZnO- and GaN-based ferromagnetic materials and their integration with commonly used III-Vs remain problematic, and the predictions have not been verified unambiguously. Room temperature ferromagnetism has been recently observed in $TiO_2$ containing Co, [11] but the integration of $TiO_2$ with conventional semiconductors for spintronic applications presents significant difficulties.

One approach to increasing $T_C$ and the quality of III-V ferromagnetic semiconductors is to incorporate Mn into the host materials in the form of digital alloys; GaAs/Mn digital alloys have been explored recently. [12, 13] The incorporation of monolayers (MLs) of materials or dopants has been widely used for fabricating digital alloys of various types, and also for so-called δ-doping of semiconductors. In the present case, Mn is expected to substitute for the group III element — in the dilute situation substitutional Mn on the group III site is an acceptor. This approach would lead to submonolayers of MnAs in GaAs or InAs, or MnSb in GaSb. Because only half a ML of Mn can be deposited with this technique, [12, 13] lateral 2D islands of MnAs or MnSb (depending on the host material) can be expected in addition to randomly distributed Mn ions within the Mn-containing layers. Some diffusion of Mn into adjacent layers of the host lattice is also expected. The advantage of such structures is that the carriers in a digital alloy can interact with both the magnetic and the semiconducting components of the structures, as is the case for electronic properties in nonmagnetic digital alloys. This contrasts with the case of 3D precipitates embedded in III-V structures, for which there appears to be no carrier interaction with ferromagnetism. [14]

The mechanism(s) of ferromagnetism for GaAs/Mn digital alloys remain unclear, [15] partly because of the large range of hole concentrations and the observation of both metallic and activated electrical transport with similar $T_C$'s. [12, 13] The absence of established theories and the likelihood of lateral 2D MnAs or MnSb islands in the digital alloys under investigation suggests that a comparison of $T_C$ for bulk MnAs and MnSb might be useful as a guide to understanding the qualitative behavior and suggesting fruitful avenues to explore in the search for higher $T_C$. The Curie temperature for MnAs is 310 K, while that for MnSb is 580 K. This suggests that GaSb/Mn might be an interesting candidate for increased $T_C$.

Recently, random alloys of GaMnSb have been grown by molecular beam epitaxy (MBE) and their structural, magnetic and electrical properties characterized. [5, 14] Two types of samples were studied, high-temperature-grown ($T_{growth}$ > 550°C) GaMnSb samples that contain 3D MnSb precipitates, and low-temperature (LT)-grown GaMnSb samples that do not show precipitates. As is the case for GaMnAs, low-temperature growth is necessary to avoid MnSb precipitation. Samples grown at high temperatures with 3D MnSb precipitates are characterized by nearly temperature independent hysteresis loops between liquid helium temperature and room temperature. [14] There is no observable anomalous Hall effect (AHE) in the same temperature range, indicating no measurable interaction between the carriers and the ferromagnetic precipitates. The exchange interaction between spins of itinerant carriers and the localized electrons in Mn ions is of critical importance because it is the basic mechanism for achieving carrier spin polarization. In contrast, in the GaMnSb samples grown at low temperature without precipitates, a strong AHE was observed below 25 K with a negative AHE coefficient; [5] this was taken to be evidence of ferromagnetism, similar to the case for LT-grown GaMnAs. From magnetotransport data the highest $T_C$ reported for GaMnSb random alloys is around 25 K, much lower than that for GaMnAs. [5]

Rather than GaMnSb random alloys, in this work we examine GaSb/Mn digital alloys grown by MBE on (100) GaAs substrates. Because of the large lattice mismatch between GaSb and GaAs (7.5%), the digital alloys were grown on GaSb buffer layers (nominally 500 nm). The growth was monitored with reflection high-energy electron diffraction (RHEED), which is a useful tool for observing the incipient formation of 3D MnSb precipitates during growth. The GaSb/Mn digital alloys consist of 50 periods of 0.5 ML Mn layers separated by GaSb layers of various thicknesses. For all the samples discussed in this report, there was no indication of 3D precipitate formation in the RHEED patterns. This, however, does not preclude the formation of such precipitates during subsequent growth before cooling down to room temperature at the end of the growth sequence. For electrical measurements, $p^+$ contacts were made in a van der Pauw configuration by Au/Zn/Au metallization and subsequent diffusion at 250°C.

We characterized the samples with superconducting quantum interference device (SQUID) magnetometry, transmission electron microscopy (TEM), magnetic force microscopy, and magnetotransport measurements in the van der Pauw configuration. This combination of measurements allowed us to elucidate the magnetic and the structural properties, as well as the interaction between the Mn spins and the itinerant carriers. The TEM image of one of the digital alloys, consisting of 0.5 ML of Mn and 12 MLs of GaSb as the repeat unit, is shown in Fig. 1. The image clearly shows the well-resolved 2D Mn-containing layers (dark) and the GaSb spacer layers (light), and indicates good structural quality. We note that the thickness of Mn-containing layers cannot be estimated from the micrograph, because TEM is sensitive to the strain distribution, rather than the chemical composition. Importantly, there is *no indication* of 3D MnSb precipitates at this resolution. [16]

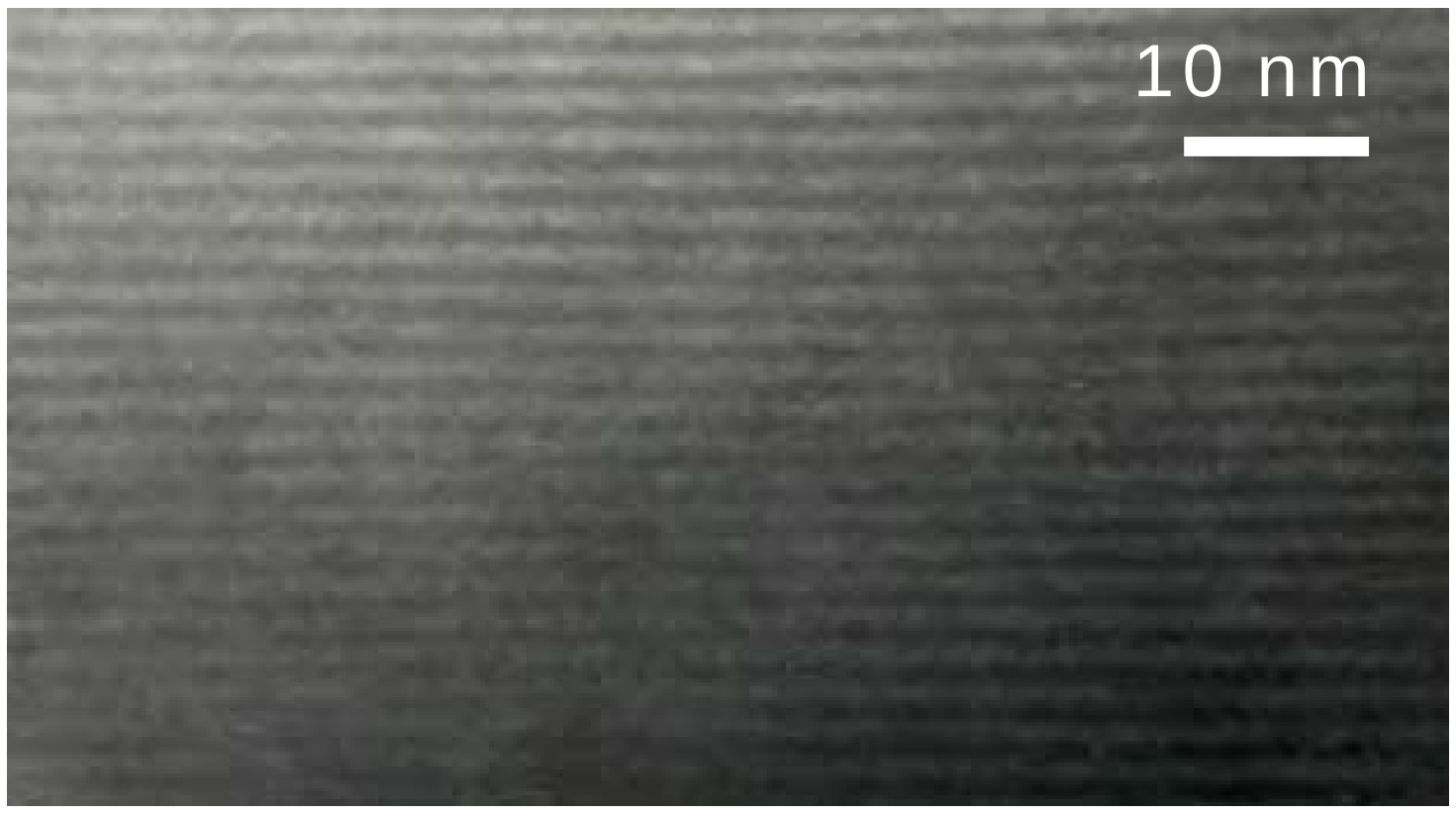

**Figure 1.** TEM image of a digital alloy with a repeat unit of Mn (0.5 ML)/GaSb (12 ML). The dark lines correspond to Mn-containing layers.

All samples showed ferromagnetism above room temperature, as indicated by hysteresis loops in the magnetization (all samples studied showed qualitatively similar magnetic behavior, and thus we show data for only one sample). The ferromagnetism is observed at temperatures up to 400 K, the upper limit of our magnetometer, and thus we can only place the lower limit that $T_C > 400$ K. The hysteresis loops show clear temperature dependence over the entire range of temperatures studied, as can be seen in Fig. 2. Note that the coercive fields at 5 K and 285 K are dramatically different, 0.01 T and 0.005 T, respectively. While ferromagnetic behavior at room temperature is also seen in GaMnAs, InMnAs and GaMnSb when there are 3D MnAs (or MnSb) precipitates, samples containing such precipitates have temperature-independent hysteresis loops in which the room temperature coercive field is nearly identical to that at 5 K. One can thus conclude from the SQUID measurements that the observed ferromagnetism in the present samples is not due 3D MnSb precipitates. Furthermore, the shapes of the hysteresis loops for the present digital alloys differ significantly (at all temperatures) from those of MnSb precipitates, [14] which can be induced in our samples by annealing at high temperatures, as discussed below.

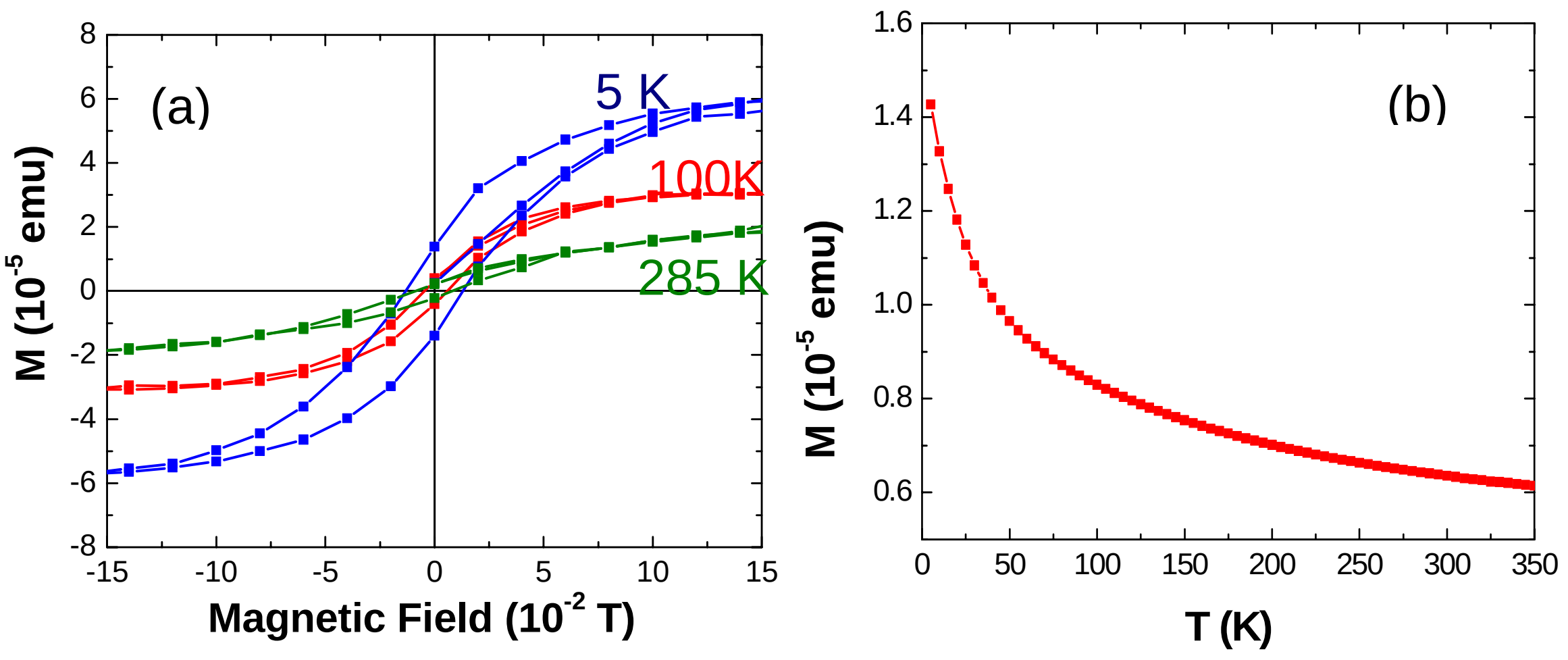

**Figure 2**. (a) Hysteresis loops of the sample shown in Fig. 1, observed with a SQUID magnetometer and (b) temperature dependence of the remanent magnetization.

Our magnetotransport measurements allow us to examine the interactions between charge carriers and magnetic ions, as has been done in previous studies of magnetic semiconductors. [5, 6, 14] All samples exhibit metallic behavior, rather than the thermally activated behavior observed in our studies of GaAs/Mn digital alloys. [13] The zero-field resistance was only weakly dependent on temperature. The carrier density estimated from the low-temperature, high-field region of the AHE curves (see Fig. 3) is approximately $3 \times 10^{13}/\mathrm{cm}^{-2}$ per Mn layer at 4.2 K, corresponding to approximately 10% of the nominal 0.5 ML Mn per Mn-containing layer. This large hole density is important for the hole-mediated exchange interaction between Mn ions.

As discussed earlier, one of the most important properties of ferromagnetic semiconductors is the interaction between itinerant carriers and localized electron spins in Mn ions. The AHE provides information about the Mn-generated internal field experienced by itinerant carriers. The AHE first decreases with temperature but remains strong up to 400 K (data are shown in Fig. 3 for 4K and 400K for simplicity) – the slope of the Hall resistance near zero-field is a measure of the magnetization, as well as strength of its coupling with the carriers. The sign of the AHE is related to the band structure of the itinerant carriers and the spin–orbit interaction. [17] We note that other possibilities, such as two carrier (electrons and holes) conduction, can also lead to behavior like that of Fig. 3. To explain the experimental results at high temperature by this mechanism, however, requires that electron conduction dominates at low fields (to explain the negative slope of $R_H$). The electrical contacts to the samples were p-type and ohmic characteristics were observed in I-V measurements over the entire temperature range. If electron conduction were to dominate, rectifying behavior should be observed in the I-V measurements. The hysteretic behavior, on the other hand (demonstrated by the open loops in the inset to Fig. 3 at 4 K and 400K), is an *unambiguous signature* of both ferromagnetism for the magnetic component and its interaction with itinerant carriers.

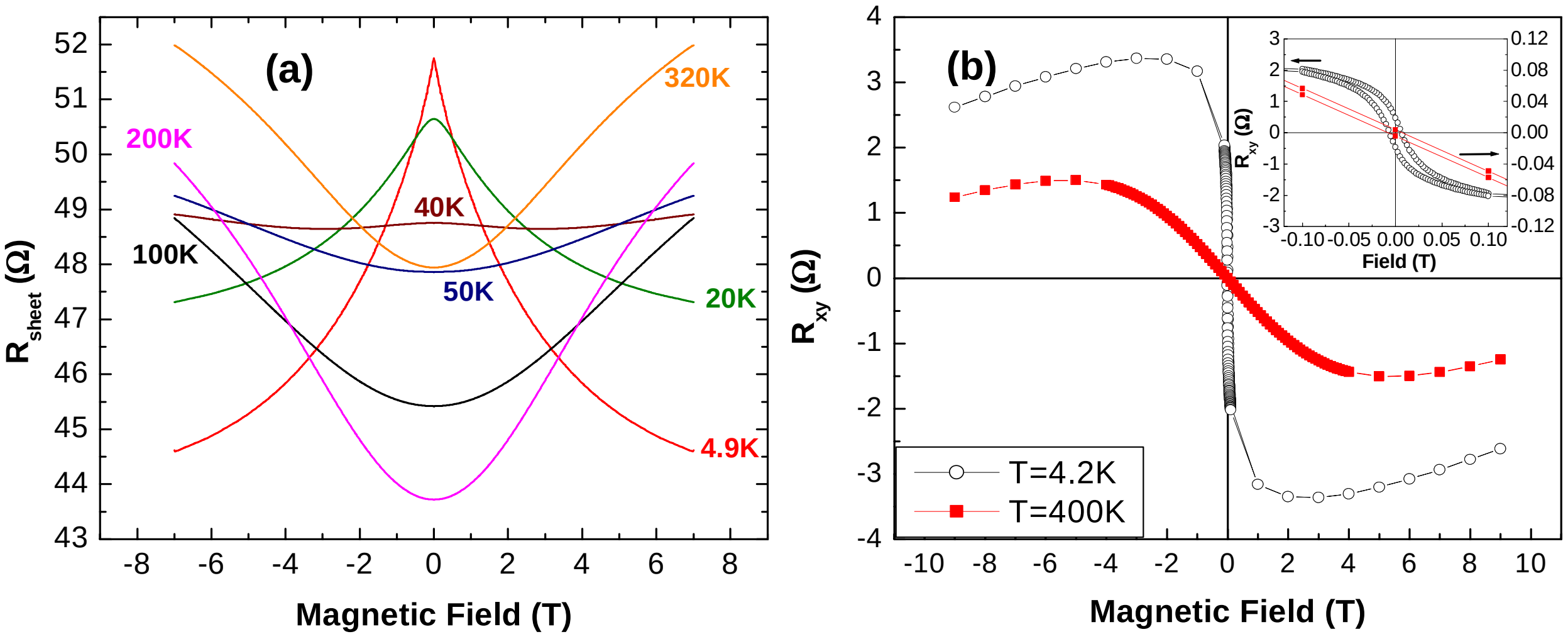

**Figure 3**. (a) The temperature dependence of the magnetoresistance and (b) the anomalous Hall effect of the sample shown in Fig. 1. The insert in (b) is the expanded plot for the low field region.

As mentioned above, previous work on GaMnSb samples containing 3D MnSb precipitates showed no clear AHE even at liquid helium temperature. [14] The AHE observed in low-temperature-grown GaMnSb random alloys with no MnSb precipitates decreased rapidly above $T_C$ (25 K), vanishing completely around 50 K. This behavior is clearly related to ferromagnetism in the random alloys. [5] The AHE coefficient is negative for the GaSb/Mn digital alloys at all temperatures, as for GaMnSb random alloys. [5] This is presently not understood. The AHE itself, even in "conventional" ferromagnetic metals, is not fully understood. [17, 18]

To further support our assertion that 3D MnSb precipitates are not the source of the observed phenomena, we have also studied the digital alloy samples annealed at high temperatures. Samples were annealed at 400°C and 500°C for 5 minutes and examined by SQUID magnetometry and atomic force microscopy/magnetic force microscopy (AFM/MFM). The combination of AFM and MFM is a powerful tool in investigating this problem. Both SQUID and MFM showed the appearance of 3D precipitates in samples annealed at 500°C. Magnetization measurements showed that the coercive field at 5 K increased from 0.01 T for the as-grown sample to 0.04 T for a sample annealed at 500°C. The shape of the hysteresis loop also became identical to that reported for MnSb precipitates. [12] Large numbers of precipitates were observed for the sample annealed at 500°C by MFM, while there was no visible change of surface structure from AFM images. Magnetic force microscopy images of the same sample before and after annealing are shown for comparison in Fig. 4. The as-grown samples appear to have magnetic domains larger than the maximum scanning area (40 μm × 40 μm) of our AFM/MFM setup, as indicated by uniform magnetic signals from the sample surface.

Temperature dependent magnetization and magnetotransport data both reveal additional complexity to the ferromagnetism in the as-grown samples. In the temperature dependence of the remanent magnetization, one finds an initial rapid drop with temperature at low temperatures, which changes to a slower decrease at higher temperatures. This change occurs between 30K and 50 K, as shown in Fig. 2. The behavior does not correspond to any of the typical power-law behaviors of magnetization seen in other thin films of ferromagnetic materials, [19] which suggests possible coexistence of more than one magnetic phase.

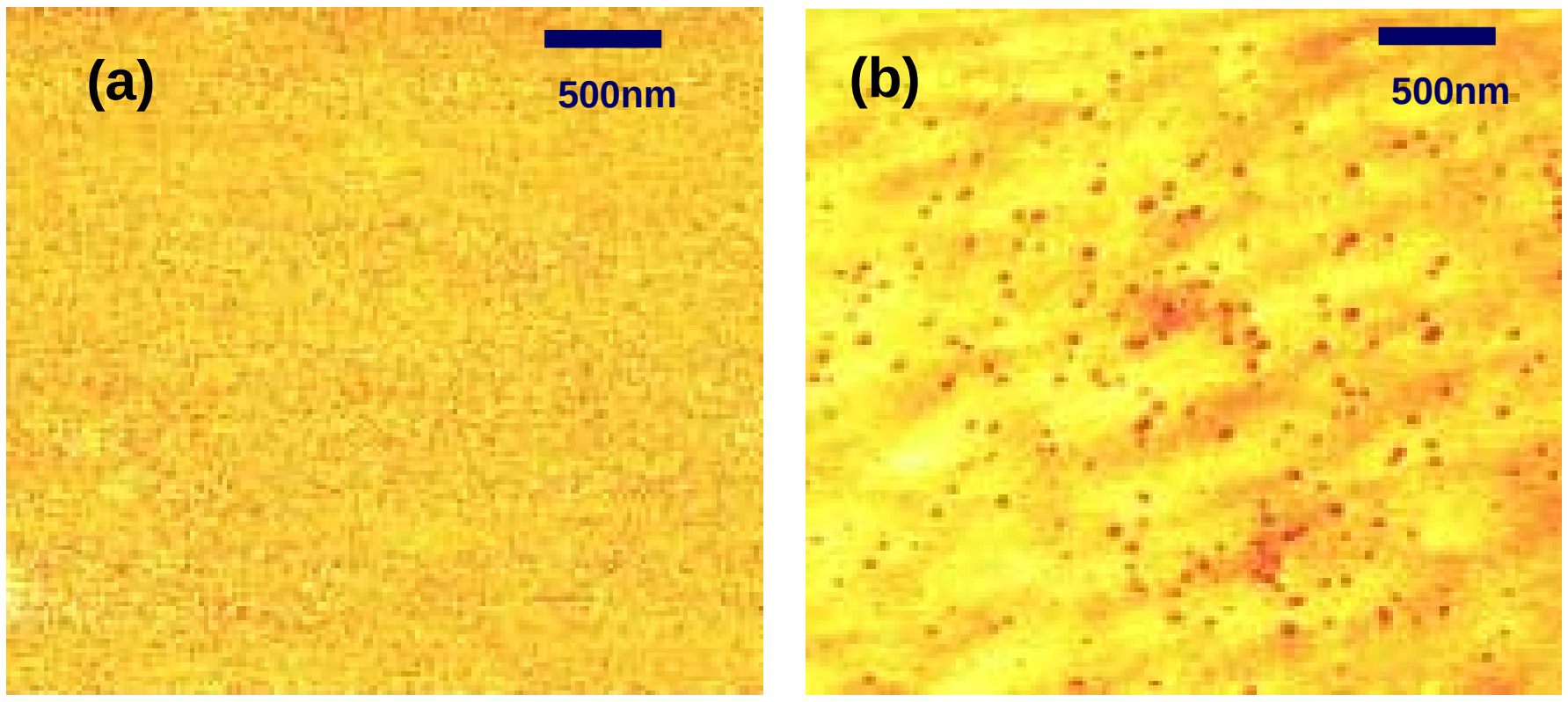

**Figure 4**. Magnetic force micrographs of a GaSb/Mn sample before (a) and after (b) annealing at 500°C. The "dots" in (b) are magnetic precipitates.

The magnetoresistance shown in Fig. 3a also changes from negative to positive with increasing temperature; this crossover takes place around 40 K. The behavior below 40 K is qualitatively the same as that for previously studied LT-grown GaMnSb random alloys, while the positive magnetoresistance above 40 K is dramatically different. [5] In the case of LT-grown GaMnSb random alloys, the negative magnetoresistance decreases quickly once the temperature is raised above $T_C$ (25 K) and the sheet resistance becomes independent of magnetic field above 50 K. Magnetoresistance is also absent in GaMnAs random alloys once the temperature is substantially above $T_C$. The significant positive magnetoresistance above 40 K (comparable to the negative magnetoresistance) in the GaSb/Mn digital alloys indicates that the interaction between holes and Mn persists to much higher temperatures, consistent with the AHE in these samples. This points to the possibility of a different magnetic phase at higher temperatures. An explanation of this positive magnetoresistance is presently lacking.

Based on the present results we suggest the following picture. The Mn-containing layers consist of 2D GaMnSb random alloys (similar to 3D GaMnSb random alloys with randomly distributed Mn ions) and 2D islands of zinc-blende MnSb. The isolated Mn portion (2D

GaMnSb random alloy) is ferromagnetic below 30-50 K, depending on the sample. This is responsible for some similarities between the GaSb/Mn digital alloys and GaMnSb random alloys at low temperatures. The observed ferromagnetism at higher temperatures is associated with the 2D MnSb islands, which may be related to the high $T_C$ in bulk MnSb (580 K). A detailed theoretical study is needed to fully understand the mechanisms involved.

In summary, we have fabricated GaSb/Mn digital alloys that exhibit ferromagnetism above room temperature. Transmission electron microscopy showed good crystal quality with well-defined 2D Mn layers separated by GaSb layers. Magnetization measurements showed that the samples are ferromagnetic above 400 K and magnetotransport measurements showed a strong AHE at 400 K. Since no AHE is observed in GaSb containing MnSb precipitates, the present studies show that the observed above-room-temperature ferromagnetism is not due to 3D MnSb precipitates. This is consistent with *in situ* RHEED measurements and *ex situ* TEM studies of these samples. The magnetization and MFM studies of annealed samples and the onset of precipitate formation with increasing annealing temperature further argue against the presence of 3D MnSb precipitates in the as-grown samples. Irrespective of the details of the mechanism, the fact that transport properties are strongly modified through *hole/Mn interaction* and hysteresis is observed in the AHE at 400 K, is very promising for future applications of such structures as spin polarizers and injectors at room temperature. This work was supported by the DARPA SPINS program through the Office of Naval Research under Grant ONR N000140010819.